\documentclass[final, numberedheadings] {aipproc}
\layoutstyle{8x11single}
\usepackage{color}  
\usepackage{amsmath} 

\newcommand{\bi}{\begin{itemize}}
\newcommand{\ei}{\end{itemize}}

\newcommand{\bt}[1]{\begin{table}[!b]\begin{tabular}{#1} \hline\hline  \\[-0.5em]}
\newcommand{\et}[2]{\hline\hline \end{tabular} \caption{#1} \label{#2} \end{table}}

\newcommand{\be}{\begin{equation}}
\newcommand{\ee}{\end{equation}}

\newcommand{\bea}{\begin{eqnarray}}
\newcommand{\eea}{\end{eqnarray}}

\newcommand{\cpt}{\ensuremath{\raise0.4ex\hbox{$\chi$}PT}}
\newcommand{\scpt}{S\ensuremath{\raise0.4ex\hbox{$\chi$}}PT}
\newcommand{\rscpt}{rS\ensuremath{\raise0.4ex\hbox{$\chi$}}PT}

\begin{document}

\title{$B$ Mixing in the Standard Model and Beyond: Lattice QCD}


\author{C.~Bernard$^1$, C.M.~Bouchard$^{2,3}$, A.X.~El-Khadra$^2$, E.D.~Freeland$^{2,4}$, E.~G\'amiz$^{3,5}$, A.S.~Kronfeld$^3$, J.~Laiho$^6$, R.S.~Van~de~Water$^7$\\
For the Fermilab Lattice and MILC Collaborations}{
address={$^1$Department of Physics, Washington University, St.~Louis, Missouri 63130, USA,\\
		$^2$Physics Department, University of Illinois, Urbana, Illinois 61801, USA,\\
		$^3$Fermi National Accelerator Laboratory, Batavia, Illinois 60510, USA,\\
		$^4$Department of Physics, Benedictine University, Lisle, Illinois, 60532, USA,\\
		$^5$CAFPE and Departamento de Fisica Teorica y del Cosmos, Universidad de Granada, Campus de Fuente Nueva, E-18002 Granada, Spain,\\
		$^6$Department of Physics and Astronomy, University of Glasgow, Glasgow, Scotland, UK,\\
		$^7$Physics Department, Brookhaven National Laboratory, Upton, New York 11973, USA}}

\begin{abstract}
We give a brief overview and progress report on our lattice QCD calculation of neutral $B$ mixing hadronic matrix elements needed for  Standard Model and Beyond the Standard Model physics.
Reference~\cite{Bmix_Lat11} contains more details and results.
\end{abstract}

\maketitle


\section{Motivation and Connection to Experiment}

Neutral-$B$ mixing is suppressed in the Standard Model, making it a process sensitive to new physics. 
The effective hamiltonian that describes neutral-$B$ mixing
$
	\mathcal{H}_{\rm eff} = \sum_{i=1}^5 C_i \mathcal{O}_i 
$
contains five operators whose matrix elements elements $\langle \bar{B}_q^0 | \mathcal{O}_i(\mu) | B_q^0 \rangle$ must be calculated with lattice QCD,
\be
	\begin{array}{l l}
	\mathcal{O}_1 = (\bar{b}^\alpha \gamma_\mu L q^\alpha )  \; (\bar{b}^\beta \gamma_\mu L q^\beta ),	&
	\mathcal{O}_4 = (\bar{b}^\alpha L q^\alpha )  \; (\bar{b}^\beta R q^\beta ), \\
	\mathcal{O}_2 = (\bar{b}^\alpha L q^\alpha )  \; (\bar{b}^\beta L q^\beta ),	&
	\mathcal{O}_5 = (\bar{b}^\alpha L q^\beta )  \; (\bar{b}^\beta R q^\alpha ),	\\
	\mathcal{O}_3 = (\bar{b}^\alpha L q^\beta )  \; (\bar{b}^\beta L q^\alpha ),
	\end{array}
\ee
where $\alpha$ and $\beta$ are color indices.
The first three operators are needed to describe Standard Model processes.
The remaining two, $\mathcal{O}_4$ and $\mathcal{O}_5$, appear in Beyond the Standard Model physics.


In the Standard Model, the meson mass difference can be expressed as
\be
	\Delta M_q = 
		\left( \frac{G_F^2 M_W^2 S_0}{4 \pi^2 M_{B_q}} \right)  \eta_B(\mu)  
		|V_{tb} V_{tq}^*|^2 
		\langle \bar{B}_q^0 | \mathcal{O}_1(\mu) | B_q^0 \rangle .  \label{eq:massdiff}
\ee
Measurements of $\Delta M_q$ have sub-percent errors~\cite{Abulencia:2006ze, Nakamura:2010zzi},
so our ability to to constrain the CKM matrix contribution $|V_{tb} V_{tq}^*|$ is limited by how precisely we know $\langle \bar{B}_q^0 | \mathcal{O}_1(\mu) | B_q^0 \rangle$.
Also of interest is the SU(3)-breaking ratio $\xi$, defined by
\be
	\frac{\Delta M_s}{\Delta M_d} 
	=  \left| \frac{V_{ts}}{V_{td}} \right|^2 	 \frac{M_{B_d}}{M_{B_s}}	\frac{\langle \bar{B}_s^0 | \mathcal{O}_1(\mu) | B_s^0 \rangle}  {\langle \bar{B}_d^0 | \mathcal{O}_1(\mu) | B_d^0 \rangle} 
	\equiv \left| \frac{V_{ts}}{V_{td}} \right|^2  \frac{M_{B_s}}{M_{B_d}}\xi^2.
\ee
$\xi$ is useful since some (lattice QCD) errors cancel in the ratio of matrix elements.
%
%
Current lattice QCD calculations have errors of just under 7\% on the square root of $\langle B_q^0 | \mathcal{O}_1(\mu) | \overline{B}_q^0 \rangle$ and just under 3\% on $\xi$~\cite{Gamiz:2009ku, Evans:2009du, Albertus:2010nm}.

Including Beyond the Standard Model effects results in a generalized version of Eq.~(\ref{eq:massdiff}),
\be
	\Delta M_q = \sum_{i = 1}^5  C_i(\mu) \langle B_q^0 | \mathcal{O}_i(\mu) | \overline{B}_q^0 \rangle .	\label{eq:massdiff_BSM}
\ee
Combined with lattice QCD results for all five matrix elements, 
Eq.~(\ref{eq:massdiff_BSM}) can be used to check that a Beyond the Standard Model prediction is consistent with experiment.
The full set of matrix elements was calculated by Ref.~\cite{Becirevic:2001xt} in 2001 using the quenched approximation.  
Advances since then allow for an improved calculation.  

The lifetime difference, in the Standard Model, can be written schematically as~\cite{Lenz:2006hd, Beneke:1996gn}
\be
	\Delta \Gamma_q = f_{B_q}^2 \left[ G_1 B_{1,q} + G_3 B_{3,q}    \right] \cos \phi_q + O(1/m_b, \alpha_s)  ,  \label{eq:deltagamma_th}
\ee
where $B_i$ are bag parameters defined via
$\langle \bar{B}_q^0 | \mathcal{O}_i(\mu) | B_q^0 \rangle   \propto f^2_{B_q} B_i(\mu) $,
the $G_i$ are calculated perturbatively, and $\phi_q$ is the CP-violating phase.
In Eq.~(\ref{eq:deltagamma_th}), $\Delta \Gamma$ is dominated by the $ \mathcal{O}_1$ contribution, resulting in a reduction of hadronic uncertainty in the ratio $\Delta \Gamma / \Delta M$.
Nevertheless, a calculation of the ratio $\langle \bar{B}_q^0 | \mathcal{O}_3(\mu) | B_q^0 \rangle / \langle \bar{B}_q^0 | \mathcal{O}_1(\mu) | B_q^0 \rangle$ on the lattice can be useful.
In addition, Eq.~(\ref{eq:deltagamma_th}) constrains the expected behavior of $\Delta \Gamma$ versus $\phi$~\cite{Lenz:2006hd, CDF9787_D05928_Jpsiphi}, and a calculation of certain combinations of the Standard Model operators should reduce the error in that relationship.

\section{The Calculation in Lattice QCD and a Discussion of Errors}

In these proceedings, we provide an overview of our error analysis; a more detailed description of the calculation can be found in Ref.~\cite{Bmix_Lat11}.
The physics of a quantum field theory can be obtained from correlation functions which, when expressed in path integral form, can be evaluated numerically on a discretized volume of space-time (the lattice).
These correlation functions can also  be written as functions of matrix elements.
%
%
There are a number of sources of error in a lattice QCD calculation.  
We review these and highlight improvements we have made over our previous calculation~\cite{Evans:2009du}.

A lattice QCD calculation obviously introduces an error due to discretization.
Guided by theory, we can extrapolate from a finite lattice spacing to the continuum and also quantify the discretization errors\footnote{Light-quark discretization errors are folded into the chiral extrapolation; heavy-quark discretization errors are often quantified separately.}.
Our current lattice QCD calculation of the $B$-mixing matrix elements includes two lattice spacings, 0.12 and 0.09~fm, included in our previous calculation as well as a third, smaller spacing of 0.06~fm.  In addition, we plan to anchor our continuum extrapolation with a fourth spacing of 0.04~fm.  In this way, we can gain better control over discretization errors.

Numerical integration involves ensemble averages over gauge (vacuum) configurations.  
The number of gauge configurations available for each ensemble has increased by a factor of 2 -- 4 for most of our data allowing for improved statistical errors.

It is computationally expensive to calculate with up and down quarks at their physical mass, so we calculate with light quarks that are ``too heavy''.  We then use chiral perturbation theory to extrapolate to the physical quark masses.
Because the light-mass pions of \cpt are the component affected by the finite volume in the simulations, the finite-volume error is folded into the chiral extrapolation.
The range of ``light'' quark masses we use spans from the strange-quark mass $m_{\rm s}$ down to $0.1 m_{\rm s}$, with one ensemble at $0.05 m_{\rm s}$.  Such a range allows for good control over the chiral extrapolation and the decrease in statistical errors mentioned above improves these extrapolations.

Lattice calculations also have errors from inputs: the physical scale of the lattice spacing and the bare quark masses.
Each of these is determined by a separate lattice QCD calculation.
Once determined, these inputs can be used for a suite of calculations of which $B$-mixing matrix elements are just one part.  
Details on the determination of these inputs can be found in Refs.~\cite{Bazavov:2009bb, Bernard:2010fr}.
Finally, operators must be matched to the continuum and renormalized.  For this we use one-loop perturbation theory resulting in a  perturbation-theory truncation error.

In the calculations discussed here, we generate numerical data for two-point and three-point correlation functions and fit them simultaneously for each meson (combination of heavy and light quarks). 
Once we have renormalized the matrix elements, we use SU(3), partially-quenched, staggered \cpt 
for the extrapolation of each matrix element and ratios of matrix elements.

Table~\ref{tbl:error} compares the error from our previous analysis to the error we expect to have for the calculation described here, based on our preliminary analysis.


\bt{l   c c c}[!t]
	source							&  2008-09			& expected 	\\	
	\hline				
	Inputs:	\\
	$\quad$scale ($r_1$)						& 3.0 			&  1.1 		\\  
	$\quad$sea, valence quark masses (mis)tuning			& 0.3 			& 0.3	\\
	$\quad$b-quark mass (mis)tuning						& 1.1 			&  $\sim 1$		\\  [0.5em]  
	statistical							& 2.7			& $\sim 1$	\\ [0.5em]   
	heavy-quark discretization			& 2.0			& $\sim$ 1.2	\\ [0.5em]  
	$\chi$PT + light quark discretization + finite volume		& 0.7 			& $\le0.7$	\\[0.5em] 
	matching, renormalization (1-loop PT)	& $\sim$ 4 		& $\sim$ 2.5	\\ 
	\hline
	total								& 6.2\%			& $\sim 3.4$\% \\
\et{Error budget on $\beta_{B_s} = \sqrt{f_{B_s}^2 M_{B_s} B}$ in percent.  Column ``2008-09'' is based on our calculation in Ref.~\cite{Evans:2009du}.  ``Expected'' is our projection based on the preliminary calculation presented here.}{tbl:error}

\section{Conclusion and Summary}

We have a good start on a large-data-set lattice QCD calculation of the matrix elements that describe neutral-$B$ mixing.
Our calculation will cover the operators needed for both Standard Model and Beyond the Standard Model physics.
For the Standard Model matrix elements, we expect to halve the error on current, published calculations.
In the Beyond the Standard Model case, this will be the first full-QCD (unquenched) calculation and the first update in ten years.


\begin{theacknowledgments}
%
Computations for this work were carried out with resources provided by
the USQCD Collaboration, the Argonne Leadership Computing Facility,
the National Energy Research Scientific Computing Center, and the 
Los Alamos National Laboratory, which are funded by the Office of Science of the
U.S. Department of Energy; 
and with resources provided by the National Institute for Computational Science, 
the Pittsburgh Supercomputer Center, the San Diego Supercomputer Center, 
and the Texas Advanced Computing Center, 
which are funded through the National Science Foundation's Teragrid/XSEDE Program.
This work was supported in part by the U.S. Department of Energy under 
Grants No.~DE-FG02-91ER40677 (C.M.B, E.D.F., E.G., A.X.K.) and No.~DE-FG02-91ER40628 (C.B.),
by the URA Visiting Scholars' program (C.M.B., E.G.), and 
by the Fermilab Fellowship in Theoretical Physics (C.M.B.).
This manuscript has been co-authored by employees of Brookhaven Science
Associates, LLC, under Contract No. DE-AC02-98CH10886 with the 
U.S. Department of Energy.
R.S.V. acknowledges support from BNL via the Goldhaber Distinguished Fellowship.
 Fermilab is operated by Fermi Research Alliance, LLC, under Contract
No.~DE-AC02-07CH11359 with the United States Department of Energy.
\end{theacknowledgments}
 
\bibliographystyle{aipproc}   

\end{document}